\documentclass[%
 reprint,
superscriptaddress,
 amsmath,amssymb,
 aps,
]{revtex4-1}

\usepackage{graphicx}% Include figure files
\usepackage{dcolumn}% Align table columns on decimal point
\usepackage{bm}% bold math
\usepackage[colorlinks= true, linkcolor=blue, citecolor=blue, urlcolor=blue]{hyperref}% add hypertext capabilities
\usepackage{lipsum}
\usepackage{bbold}
\usepackage{mathtools}
\usepackage[normalem]{ulem}
\usepackage{hyperref}
\usepackage{enumitem}

\def\bea{\begin{eqnarray}}
\def\eea{\end{eqnarray}}

\newcommand{\aref}[1]{Appendix \ref{#1}}%

\usepackage{xcolor}

\newcommand{\eref}[1]{Eq.~(\ref{#1})}
\newcommand{\fref}[1]{Fig.~\ref{#1}} 

\begin{document}

%\preprint{APS/123-QED}

%\title{Expediting classical first passage resetting with stochastic returns}
%\title{Expediting drift-diffusive search process facilitated by stochastic resetting with stochastic return}
%\title{Expediting drift-diffusive resetting search process by non-instantaneous stochastic returns}
\title{Drift-diffusive resetting search process with stochastic returns: speed-up beyond optimal instantaneous return}
%\title{Fluctuations in return time can expedite classical first passage under resetting}

\author{Arup Biswas}
\email{arupb@imsc.res.in}
\affiliation{The Institute of Mathematical Sciences, CIT Campus, Taramani, Chennai 600113, India \& Homi Bhabha National Institute, Training School Complex, Anushakti Nagar, Mumbai 400094, India}
\author{Ashutosh Dubey}
\email{ashutoshrd@imsc.res.in}
\affiliation{The Institute of Mathematical Sciences, CIT Campus, Taramani, Chennai 600113, India \& Homi Bhabha National Institute, Training School Complex, Anushakti Nagar, Mumbai 400094, India}
\author{Anupam Kundu}
\email{anupam.kundu@icts.res.in}
\affiliation{International Centre for Theoretical Sciences, TIFR, Bangalore, India}

\author{Arnab Pal}
\email{arnabpal@imsc.res.in}
\affiliation{The Institute of Mathematical Sciences, CIT Campus, Taramani, Chennai 600113, India \& Homi Bhabha National Institute, Training School Complex, Anushakti Nagar, Mumbai 400094, India}

%\date{\today}

\begin{abstract}
Stochastic resetting has emerged as a useful strategy to reduce the completion time for a broad class of first passage processes. In the canonical setup, one intermittently resets a given system to its initial configuration only to start afresh and continue evolving in time until the target goal is met. This is, however, an instantaneous process and thus less feasible for any practical purposes. A crucial generalization in this regard is to consider a finite-time return process which has significant ramifications to the first passage properties. Intriguingly, it has recently been shown that for diffusive search processes, returning in finite but stochastic time can gain significant speed-up over the instantaneous resetting process. Unlike diffusion which has a diverging mean completion time, in this paper, we ask whether this phenomena can also be observed for a first passage process with finite mean completion time. To this end, we explore the set-up of a classical drift-diffusive search process in one dimension with stochastic resetting and further assume that the return phase is modulated by a potential $U(x)=\lambda |x|$ with $\lambda>0$. For this process, we compute the mean first passage time exactly and underpin its characteristics with respect to the resetting rate and potential strength. We find a unified phase space that allows us to explore and identify the system parameter regions where stochastic return supersedes over both the underlying process and the process under instantaneous resetting. Furthermore and quite interestingly, we find that for a range of parameters the mean completion time under stochastic return protocol can be reduced further than the \textit{optimally restarted} instantaneous processes. We thus believe that resetting with stochastic returns can serve as a better optimization strategy owing to its dominance over classical first passage under resetting.
\end{abstract}

\pacs{Valid PACS appear here}% PACS, the Physics and Astronomy
\maketitle
%\onecolumngrid

\section{Introduction}
Search processes are ubiquitous in nature and appear in diverse contexts -- for example,  a pigeon searching for its nest \cite{mora2004magnetoreception,wallraff2001navigation,walcott1996pigeon}, a computer algorithm searching for the optimal solution \cite{luby1993optimal}, drones for locating some specified targets \cite{mishra2020drone}, or an enzyme that searches for a particular substrate on DNA to bind \cite{chou2014first}. 
 In the context of statistical physics, target finding processes are  popularly known as first passage (FP) processes. The time to complete a FP process is known as the first passage time (FPT) \cite{redner2001}. A major aspect of a FP process is concerned with finding optimal search strategies to expedite and complete the process in a minimal time span. Animals have adapted several search strategies to find their nest or food \cite{bartumeus2005animal,o1990search,he2009group}. At the cellular level, the idea of facilitated diffusion has been accepted to answer significantly short time scales involved in chemical kinetics \cite{mirny2009protein,benichou2011facilitated}. From a more statistical mechanics viewpoint, intermittent search strategies have been explored a lot as a potential way for facilitating FP processes \cite{benichou2011intermittent,lomholt2008levy}. The quest for better and better search strategies still goes on.

Nearly a decade ago, Evan-Majumdar in their seminal work \cite{evans_diffusion_2011} reintroduced the idea of stochastic resetting as one of such search strategies. Here, one intermittently stops and restarts the search process back from where it originally started. Although seemingly counter-intuitive, resetting has been shown to immensely improve the performance of a search process. A panorama of studies followed after that to investigate the effect of stochastic resetting in a variety of systems \cite{pal2015diffusion,gupta2014fluctuating,huang2021random,chen2022first,jolakoski2022fate,kusmierz2014first,evans2018run,de2022optimal,mercado2020intermittent,majumdar2020extreme,magoni2020ising,pal2022inspection,basu2019symmetric,bonomo2021first,singh2022first,pal_first_2017,pal2016diffusion,kumar2023universal,olsen2023thermodynamic,mori2023entropy,bonomo2021mitigating,biswas2023rate,ahmad2019first,bonomo2021first,garcia2023optimal,ray2019peclet,bodrova_continuous-time_2020,ray2020space,reuveni_optimal_2016,blumer2024combining,bressloff2021accumulation,bressloff2021drift,kundu2024preface,yin2023restart,stanislavsky2023confined,reuveni2014role}.
% {\color{red}ak: we can cite the preface of the special issue on resetting in this group} \textcolor{magenta}{AB: Done.}
We refer to \cite{evans_stochastic_2020,pal2023random,nagar2023stochastic} for a detailed review of the subject.  Notwithstanding the simplicity of the model, the major hindrance turns out to be its practical applicability as it assumes a zero time teleportation of the searcher to a resetting location. 
% This immediately violates the universal speed limit obtained by Einstein that no physical object can propagate faster than the speed of light \cite{resnick1991introduction}. 
Any return event of the searcher to its starting position must take a finite amount of time. Previous studies have made such attempt either by adding an  overhead time after each reset \cite{evans2018effects,roy2024queues,garcia2024stochastic,reuveni2014role} or making the return process space-time correlated \cite{radice2022diffusion,radice2021one,pal_search_2020,gupta2020stochastic,pal2019invariants,bodrova2020brownian,maso2019transport,bressloff2020search,stanislavsky2022subdiffusive,sunil2023cost,bodrova2020brownian}. Since the return motion of the searcher was \textit{deterministic} and was always directed to the resetting location, such protocols have always contributed addition time penalty to the FPT in comparison to the classical instantaneous return processes.

In contrast, in a recent work \cite{biswas2023stochasticity} it has been shown that non-instantaneous resetting with \textit{stochastic} returns can actually supersede instantaneous return under certain conditions. Though looking counter-intuitive at first glance this can indeed be achieved by leveraging a stochastic component in the searcher's motion during return. When the searcher returns home following some random motion, there is always a chance that it finds the target. This sometime may make the search process more efficient as target detection is conducted both during search and return phase. Notice that this was not possible for the instantaneous return protocol as there the searcher is guaranteed to reach the origin in zero time. Consequently, there is no target detection probability during return. It has to restart the whole search process once a resetting event occurs. Despite its target detection probability during the return phase, one can not ignore the fact that stochastic return consumes a finite time whenever the searcher reaches the origin after the return phase. A universal criterion was found in \cite{biswas2023stochasticity}, which can ensure that stochastic return can indeed be advantageous over instantaneous resetting. For a freely diffusive Brownian particle in one dimension searching for a target, the return was accomplished by turning on a potential trap around the origin. A parameter space was found in terms of the resetting rate and the return potential strength, where the mean first passage time (MFPT) with stochastic return becomes smaller than the MFPT of instantaneous return. 

The MFPT of a free diffusive particle is infinite due to its heavy-tailed distribution \cite{feller1991introduction}. Thus any finite amount of resetting helps in making the MFPT a finite quantity by cutting off the diverging trajectories going far away from the target. But what if the underlying reset-free process itself has a finite MFPT? How resetting is supposed to perform there? This question has been answered in \cite{pal_first_2017} where it was shown that whenever the ratio of fluctuation to mean of FPT, known as the coefficient of variation ($CV$), is greater than unity, resetting with the instantaneous return is guaranteed to help. Certainly, there will be regions where resetting with instantaneous return can not be beneficial. This leaves us to wonder, whether resetting with stochastic return may come to be an advantageous strategy there. Moreover, in the regions where instantaneous resetting helps in reducing the MFPT compared to the underlying process, can stochastic return do even better? Can one find a parameter space where stochastic return proves to be fruitful in not only decreasing the MFPT compared to that of instantaneous return but also compared to the MFPT of the underlying process? Understanding these aspects is the central goal of this paper. The simplest case where MFPT is rendered finite is when a drift is added to the Brownian particle towards the target \cite{redner2001}. The drift-diffusion process also bears a close resemblance to modelling several real-life systems. Some of them include the motion of a particle under the flow field of fluid \cite{risken1985fokker}, charge transport \cite{jacoboni1977review}, neuron firing \cite{capocelli1971diffusion}, price of stocks \cite{laurent2020volatility} etc. In this paper, we consider the drift-diffusive search process in one dimension with intermittent stochastic return to the origin. The return is modulated by turning on a potential trap around the origin. We analyze the mean first passage time of the particle to reach the target. Our main focus will be to probe its parameter space and find criteria for stochastic return to be better than the underlying process as well as the instantaneous return case. 

The paper is organized as follows: In section \ref{sec2} we illustrate our drift-diffusion model in detail and find the MFPT for any arbitrary resetting time distribution. In section \ref{sec3} we take the explicit case when the resetting times are drawn from an exponential distribution. In section \ref{sec3a} we find the criterion for resetting with stochastic return to provide an advantage over the underlying process. In section \ref{sec3b} we find the criterion where stochastic return can be more beneficial than the instantaneous return. In section \ref{sec4} we merge both these criteria to have a unified parameter space from where one can identify the utility of stochastic return for any given set of parameters. Finally, in section \ref{seciv} we find the parameter space where the MFPT obtained from stochastic return protocol can be further minimized than the lowest MFPT obtained from optimal instantaneous return protocol.

%%%%%%%%%%%%%%%%%%%%%%%%%%%%%%%%%%%%%%%%%%%%%%%%%%%%%%%%%%%%%%%%%%%%%%%%%%%%%%%%%%%%%%%%%%%%%%%%%%%%%%%%%%%%%%%%%%%%%%%%%%%%%%%%%%%%%%%%%%%%
\begin{figure}
    \centering

    \includegraphics[width=8cm]{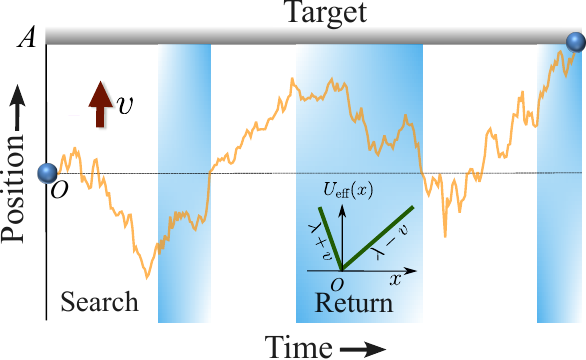}
    \caption{Drift-diffusive search process with resetting. A drift of magnitude $v$ (red arrow) always acts on the particle, biasing its motion towards the target.  The resetting is modulated by turning on a potential $U(x)=\lambda |x|$ with $\lambda>0$ at random times (shown by the blue strips). This potential causes an additional drift of magnitude $\lambda$ towards the origin. Overall an effective potential $U_{\text{eff}}(x)$ as in \eref{ueff}  acts on the particle during the return.  After the particle reaches the origin the potential is turned off marking the completion of one resetting event. The search process can end in two ways: either the target is found in the search phase (potential off) or in the return phase (potential on).}
    \label{fig1}
\end{figure}
%%%%%%%%%%%%%%%%%%%%%%%%%%%%%%%%%%%%%%%%%%%%%%%%%%%%%%%%%%%%%%%%%%%%%%%%%%%%%%%%%%%%%%%%%%%%%%%%%%%%%%%%%%%%%%%%%%%%%%%%%%%%%%%%%%%%%%%%%%%%%%%%%%%%%%%%%%%%%%%%%%%%%%

\section{Drift-diffusive search process in one dimension}
\label{sec2}
Consider a diffusive Brownian particle in one dimension (1-d) which starts its motion from $x=0$. There is a constant drift of magnitude $v>0$ that acts on the particle. This drift biases the particle's motion towards the target at $x=L>0$. Once the particle reaches the target at $x=L$ the motion is completed and we note the corresponding first passage time denoted by the random variable $T$. This problem is well studied in literature \cite{redner2001}. For completeness, we provide small derivation of the first passage probabilities in \aref{appa}.

In addition to the free drift-diffusive motion, now at random times $R$ generated from a distribution $f_R(t)$, a potential $U(x)$, centred at the origin, is turned on. This potential tries to bring the particle back towards the origin. The moment the particle reaches the origin the potential is switched off marking the completion of one resetting event. After that, it follows again follows underlying drift-diffusive motion. The entire process is repeated until and unless the target is found. The search can end in two ways: either the particle reaches the target (at $x=L$) in the search phase \textit{i.e.} while the potential is off or in the return phase \textit{i.e.} when the potential is turned on. Our aim is to analyse the MFPT for this system.

For analytical simplicity, we take the simplest form of the return potential which is a  linear one given by
\begin{align}
    U(x)=\lambda |x|.
\end{align}
This form of potential imposes another drift $\lambda>0$ on the particle on top of the underlying drift velocity $v$ but now towards the origin (in contrast to the drift towards the target) during the return phase. Note that, due to the constant nature of the drift velocity the return motion is now also a drift-diffusive process similar to the underlying motion (search phase). However, the direction and magnitude are now changed. For $x>0$, the net drift now becomes
\begin{align}
    \lambda_+=\lambda-v,
\end{align} towards the origin (assuming $\lambda>v$). Quite evidently, when $\lambda<v$ the net drift acts towards the target in the region $x>0$. On the other hand for $x<0$ the drift always is
\begin{align}
    \lambda_-=\lambda+v,
\end{align} towards the target (or origin, as both fall on the same side). In another way, one can construct an effective potential $U_{\text{eff}}(x)$ as below
\begin{align}
\begin{array}{l}
U_{\text{eff}}(x)=\left\{ \begin{array}{lll}
\lambda_+x \hspace{3cm} &\text{if }x>0,\\
 -\lambda_-x\hspace{0.6cm} &\text{if }x<0,  \end{array}\right.\text{ }\end{array}
\label{ueff} 
\end{align}
which acts on the particle during the return phase. \fref{fig1} depicts a trajectory of the particle schematically. In summary, each resetting event now has two major caveats: first, as the particle reaches the origin in a stochastic fashion, it consumes a finite time and secondly, the particle has a finite probability of reaching the target during the return. In the limit $\lambda\to\infty$ one however, expects to recover the results for the classical instantaneous return for a drift-diffusive system \cite{ray2019peclet,pal_search_2020}.

\subsection*{Mean first passage time}

Let us denote by $T_R$ the random variable associated with the first passage time of the entire process (search + return phase) 
% \textcolor{magenta}{AB: I used `search' in the schematic. Thus used `search' rather than `exploration' in all the places. Just to be consistent with the terminology.}{\color{red} AK: I think it is better to write exploration as we are emphasising that even during the return the particle is searching. I would suggest to change search $\to$ exploration at all relevant places. However I do not have strong feeling. I leave upto you and Arnab to decide.}  \textcolor{blue}{To keep it consistent with the short paper, let's use `search' \& `return'.}
described above. We are interested in finding the MFPT $\langle T_R \rangle$ of the overall search process. A general renewal formalism was presented in \cite{biswas2023stochasticity} for the MFPT of such a search process for any arbitrary return motion. Here we do not delve much into the formalism but present the general result for the MFPT in 1-d obtained in \cite{biswas2023stochasticity}. We simply write here 
\begin{align}
    &\langle T_R \rangle = \frac{\langle min(T,R) \rangle +\int_{-\infty}^Ldx\widetilde{G}_R(x)\langle t^{ret}(x) \rangle}{1-\int_{-\infty}^Ldx~\widetilde{G}_R(x) \epsilon^{ret}_O(x)},
    \label{mfpt-1d}
\end{align}
where, the quantity $\widetilde{G}_R(x)=\int_{0}^{\infty}dt f_R(t)G(x,t)$ is the time-integrated propagator with $G(x,t)$ being the propagator of the underlying process in the absence of resetting and $f_R(t)$ is the distribution of the resetting time $R$. Other quantities are defined as follows. Here 
\begin{align}
    \langle t^{ret}(x) \rangle=\theta(-x)\langle t_1(x) \rangle + \theta(x)\langle t_2(x) \rangle,\label{tret}
\end{align}
represents the mean return time either to the origin ($O$) or target ($A$) (see \fref{fig1}) starting the return motion from $x$. $\langle t_1(x) \rangle$ is the mean time to reach the origin during return from $x<0$ and $\langle t_2(x) \rangle$ is the unconditional mean time to reach either the target or the origin during the return from $x>0$. The quantity $\epsilon^{ret}_O(x)$, given by
\begin{align}
    \epsilon^{ret}_O(x)=\theta (-x)+ \theta(x) \epsilon_O(x),\label{ereto}
\end{align}
represents the splitting probability to reach the origin during return. $\epsilon_O(x)$ is the splitting probability to reach the origin while returning from $x>0$. Note that starting from $x<0$, the particle will definitely reach the origin before reaching the target at $L>0$. The minimum of the two random variables $T$ and $R$ is denoted by $min(T,R)$ where recall that $T$ denotes the random variable associated with the underlying first-passage time and $R$ is the random waiting time when the return phase commences since the start of the last search phase.

We would thus need the exact expressions for the unconditional MFPT and the splitting probabilities of the return motion facilitated by the potential $U_{\text{eff}}(x)$. In \aref{appa} we give detailed derivations of these quantities.  Here we just recall the main results which we shall be using in the later parts of this paper. The underlying reset-free propagator $G(x,t)$, and its Laplace transform $\widetilde{G}(x,s)=\int_{0}^{\infty}dt e^{-st}G(x,t)$, for the drift-diffusive process in presence of a target in 1-d are computed in the textbook \cite{redner2001}. It is given by
\begin{align}
    G(x,t)&=\frac{1}{\sqrt{4 \pi  D t}}\left(e^{-\frac{(x-t v)^2}{4 D t}}-e^{\frac{L v}{D}} e^{-\frac{(2 L+t v-x)^2}{4 D t}}\right), \label{prp}
\end{align}
which after Laplace transforming with respect to $t$ looks like
\begin{align}
     \widetilde{G}(x,s)&=\frac{e^{\frac{v x}{2 D}} \left(e^{-\frac{| x|  \sqrt{4 D s+v^2}}{2 D}}-e^{-\frac{(2 L-x) \sqrt{4 D s+v^2}}{2 D}}\right)}{\sqrt{4 D s+v^2}} .
    \label{prp_ls}
\end{align}
The MFPT to reach either the target at $x=L$ or the origin at $x=0$ during the return motion starting from $x>0$ is found to be
\begin{align}
     \langle t_2(x) \rangle = \frac{L(1-e^{\lambda_+ x/D}) +x(e^{\lambda_+ L/D }-1)}{\lambda_+(e^{\lambda_+ L/D}-1)}.
     \label{t2-1d}
\end{align}
The associated splitting probability to reach the target \textit{i.e.} $\epsilon_L(x)$ or to reach the origin \textit{i.e.} $\epsilon_O(x)$, is given by
\begin{align}
    \epsilon_L(x)=1-\epsilon_O(x)=\frac{1-e^{\lambda_+ x/D}}{1-e^{\lambda_+ L/D}}.\label{el}
\end{align}
When the return motion starts from $x<0$ the particle has to cross the origin first implying the probability of reaching the target is zero so that 
\begin{align}
    \epsilon_L(x)=1-\epsilon_O(x)=0, \label{e-neg}
\end{align}
while the corresponding MFPT to reach the origin is given by
\begin{align}
    \langle t_1(x) \rangle=\frac{|x|}{\lambda_-}. \label{t1-1d}
\end{align}
With these quantities in hand, one only requires the specific return (resetting) time distribution to make further progress. In what follows we shall take the set-up where resetting times are drawn from an exponential distribution. The choice of exponential resetting is primarily due to its advantage in giving nice closed-form results. However, we emphasize that the formalism presented till now can always be utilized for other resetting protocols as well.

\section{Resetting at exponential times}
\label{sec3}
Suppose the resetting/return times are drawn randomly from an exponential distribution with rate $r$, given by 
\begin{align}
    f_R(t)=re^{-rt}. \label{exp}
\end{align}
Recalling \eref{mfpt-1d}, we first proceed to find the quantity $\langle min(T,R) \rangle$. By definition, this is found to be (see \aref{appc1} for detailed derivation)
\begin{align}
      \langle min(T,R) \rangle =\int_0^\infty dt Pr(T> t)Pr(R> t),
       \label{prz}
\end{align}
with $f_T(t)=-\frac{dPr(T> t)}{dt}$ being the first passage time distribution of the underlying reset-free process. For exponential distribution, we obtain from \eref{prz}
\begin{align}
     \langle min(T,R) \rangle &= \int_0^\infty dt e^{-rt} \int_t^\infty dt'f_T(t') \nonumber \\
     &=\frac{1}{r}-\frac{1}{r}\int_0^\infty dt e^{-rt}f_T(t) \nonumber \\
     &=\frac{1-\widetilde{T}(r)}{r},
     \label{mintr}
\end{align}
where $\widetilde{T}(r)=\int_0^\infty dt e^{-rt}f_T(t)$
is the Laplace transform of $f_T(t)$. For the drift-diffusive search process, the first passage time density $f_T(t)$ and its Laplace transform $\widetilde{T}(s)$ are given by (see \aref{appb})
\begin{align}
    f_T(t)&=\frac{L e^{-\frac{(L-t v)^2}{4 D t}}}{2 \sqrt{\pi } \sqrt{D} t^{3/2}},~~\text{and}\label{fpt-dd}\\
    \widetilde{T}(s)&=e^{\frac{L \left(v-\sqrt{4 D s+v^2}\right)}{2 D}}, \label{tl-exp}
\end{align}
respectively.
The time-integrated propagator for the exponential resetting distribution is simply its Laplace transform multiplied by $r$ as shown below
\begin{align}
    \widetilde{G}_R(x)=\int_{0}^{\infty}re^{-rt}G(x,t)dt=r\widetilde{G}(x,r), \label{prp_exp}
\end{align}
where exact expression for $\widetilde{G}(x,r)$ is given in \eref{prp_ls}. 

\begin{figure*}
    \centering
    \includegraphics[width=16cm]{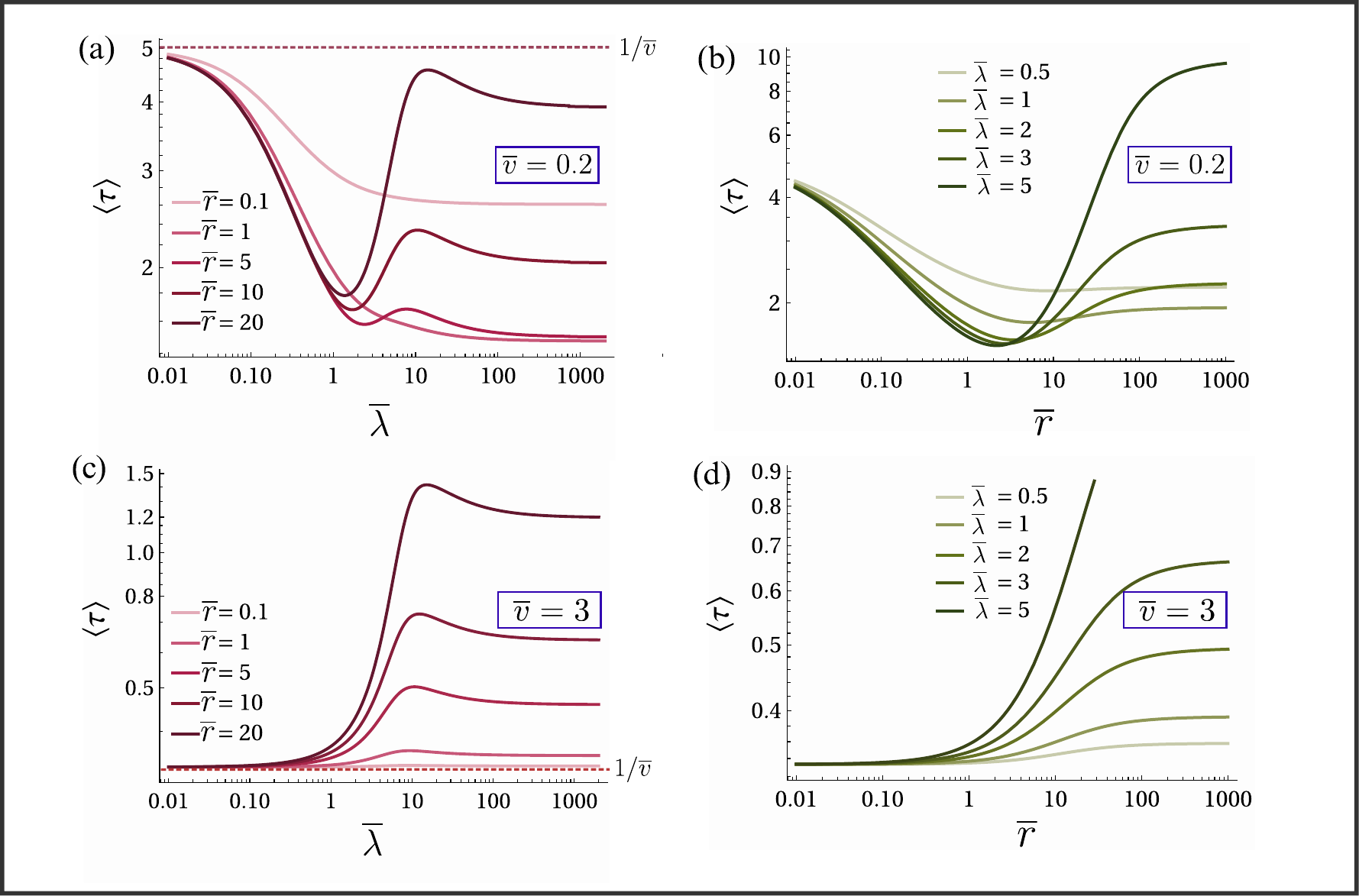}
    \caption{Variation of the MFPT $  \langle \tau(\overline{r},\overline{\lambda},\overline{v}) \rangle$ with potential strength $\overline{\lambda}$ (panel a,c) and resetting rate $\overline{r}$ (panel b,d) for exponentially distributed resetting times. Panel (a,b) are for $\overline{v}=0.2$ and panel (c,d) are shown for $\overline{v}=3$. The limit $\overline{\lambda}\to 0$ gives the MFPT of the underlying process \textit{i.e.} $\frac{1}{\overline{v}}$ shown by the dashed horizontal lines.  Note the contrasting behaviour for different values of $\overline{v}$.  For lower value of $\overline{v}$ each plot shows a non-monotonic behaviour implying the utility of resetting compared to the reset-free process. However, as $\overline{v}$ becomes higher MFPT for a finite value of $\overline{r}$ and $\overline{\lambda}$ is always greater than that of the underlying process. Also in some of the curves in panel (a) and in each of the curves in panel (c), there exists a range of $\overline{\lambda}$ where MFPT becomes lower than the instantaneous return limit ($\overline{\lambda}\to \infty$).}
    \label{fig2}
\end{figure*}

After using the results obtained from \eref{mintr} and \eref{prp_exp} to \eref{mfpt-1d}, we have
\begin{align}
    &\langle T_R (r,\lambda,v) \rangle =\frac{\frac{1-\widetilde{T}(r)}{r} +r\int_{-\infty}^Ldx\widetilde{G}(x,r)\langle t^{ret}(x) \rangle}{1-r\int_{-\infty}^Ldx~\widetilde{G}(x,r) \epsilon^{ret}_O(x)},
    \label{mfptexp}
\end{align}
where we have omitted the explicit dependence on $D$ and $L$. Recall that $\langle t^{ret}(x) \rangle,\epsilon^{ret}_O(x)$ are given by \eref{tret} and \eref{ereto} respectively. It is now a straightforward task to plug the results obtained from \eref{t2-1d}-(\ref{t1-1d}) and \eref{tl-exp} in the above equation to have the full MFPT for the drift-diffusive system with exponential resetting. It turns out that the overall MFPT can be written in a dimensionless form 
\begin{align}
    \langle \tau(\overline{r},\overline{\lambda},\overline{v}) \rangle=\frac{D}{L^2}\langle T_R (r,\lambda,v) \rangle,
\end{align}
where we introduce the scaled quantities 
\begin{align}
    \overline{r}=rL^2/D,~ \overline{\lambda}=\lambda L/D, ~ \overline{v}=v L/D.
\end{align}
The exact expression for $\langle \tau(\overline{r},\overline{\lambda},\overline{v}) \rangle$ is quite cumbersome and provided in the Mathematica file in the GitHub link \cite{key}. \fref{fig2} shows its variation with respect to the potential strength $\overline{\lambda}$ and resetting rate $\overline{r}$ for two distinct values of $\overline{v}=0.2 \text{ and } 3$.

%%%%%%%%%%%%%%%%%%%%%%%%%%%%%%%%%%%%%%%%%%%%%%%%%%%%%%%%%%%%%%%%%%%%%%%%%%%%%%%%%%%%%%%%%%%%%%%%%%%%%%%%%%%%%%%%%%%%%%%%%%%%%%%%%%%%%%%%%%%%
%%%%%%%%%%%%%%%%%%%%%%%%%%%%%%%%%%%%%%%%%%%%%%%%%%%%%%%%%%%%%%%%%%%%%%%%%%%%%%%%%%%%%%%%%%%%%%%%%%%%%%%%%%%%%%%%%%%%%%%%%%%%%%%%%%%%%%%%%%%%

The behaviour with respect to potential strength is quite intricate as can be seen from \fref{fig2}(a,c). Let us first discuss two important limiting cases of the MFPT $\langle\tau(\overline{r},\overline{\lambda},\overline{v}) \rangle$ that we shall frequently refer to in the rest of the paper.

\textbf{MFPT of the underlying (reset-free) process: }In the limit $\overline{\lambda}\to0$ no effect of resetting potential is there and the process effectively goes back to the simple reset-free drift-diffusion. One can check that in this limit the MFPT is given by
\begin{align}
      \langle \tau(\overline{r},\overline{\lambda}\to0,\overline{v}) \rangle=\frac{1}{\overline{v}}, \label{noreset}
\end{align}
which the MFPT of the underlying drift-diffusive process \cite{redner2001}. Note that this is also equivalent to setting $\overline{r}\to 0$. Both the limits correspond to the same underlying reset-free process.

\textbf{MFPT with instantaneous return: }On the other hand when $\overline{\lambda}\to\infty$ the potential strength is so high that the particle returns almost instantaneously to the origin. The resulting expression of the MFPT in this limit indeed reduces the result for the instantaneous return case
\begin{align}
      \langle \tau(\overline{r},\overline{v}) \rangle_{\text{inst}}= \langle \tau(\overline{r},\overline{\lambda}\to\infty,\overline{v}) \rangle=\frac{e^{\frac{1}{2} \left(\sqrt{\overline{v}^2+4 \overline{r}}-\overline{v}\right)}-1}{\overline{r}},\label{inst-dd}
\end{align}
as derived in \cite{ray2019peclet}. Here we have denoted the MFPT associated with instantaneous return protocol by $ \langle \tau(\overline{r},\overline{v}) \rangle_{\text{inst}}$. 

Quantification of MFPT for intermediate values of $\overline{\lambda}$ is highly non-trivial. Note that depending on the magnitude of the drift $\overline{v}$ of the underlying process the plots show two distinct kinds of behaviour. Let us first focus on \fref{fig2}(a) which is for $\overline{v}=0.2$. For lower values of resetting rates (e.g. $\overline{r}=0.1,1$), the curves monotonically go down and saturate to the instantaneous return MFPT given by \eref{inst-dd}. On the other hand, when resetting rates are high it shows non-monotonic behaviour with $\overline{\lambda}$. Particularly, for $\overline{r}=10,20$ it is evident from the plots that, for a range of $\overline{\lambda}$ \textit{MFPT can be reduced below that of both the underlying reset-free process and the instantaneous return limit}. In contrast, for the value of $\overline{v}=3$, we see from \fref{fig2}(c) that MFPT never goes below the MFPT of the underlying reset-free process for any values of $\overline{\lambda}$. However, in this case also one can find the set of $\overline{\lambda}$ values where MFPT is lower than that of instantaneous resetting.

With resetting rate $\overline{r}$ MFPT shows a non-monotonic variation for lower values of $\overline{v}$ (look at \fref{fig2}(b)). In the absence of resetting $\overline{r}\to0$ one gets 
\begin{align}
     \langle \tau(\overline{r}\to0,\overline{\lambda},\overline{v}) \rangle=\frac{1}{\overline{v}}, \label{und-process}
\end{align}
the same as in \eref{noreset} as they are physically equivalent. When resetting is too frequent \textit{i.e.} in the limit $\overline{r}\to\infty$, the potential is almost always turned on, and the MFPT resembles that of a particle in the potential  $U_{\text{eff}}(x)$, given by
\begin{align}
     \langle \tau(\overline{r}\to\infty,\overline{\lambda},\overline{v}) \rangle=\frac{\overline{v}^2-\overline{\lambda} \left(-2 e^{\overline{\lambda}-\overline{v}}+\overline{\lambda}+2\right)}{(\overline{\lambda}-\overline{v})^2 (\overline{\lambda}+\overline{v})}.
\end{align}
Once $\overline{v}$ is increased this non-monotonous behaviour goes away and MFPT starts increasing with $\overline{r}$ for any value of $\overline{\lambda}$ (see \fref{fig2}(d)).

From the above discussion, it is evident that resetting is not guaranteed to expedite the search process for any arbitrary parameters. \fref{fig2}(a,b) are the cases where it indeed helps. On the contrary, \fref{fig2}(c,d) depicts that any amount of resetting is detrimental.  In the next section, we delve deeper into this and try to find a criterion for resetting to be helpful.

\subsection{Speed up over underlying process - the \textit{CV criterion}}
\label{sec3a}
We ask: given a certain set of parameters what kind of behaviour the MFPT of our system is expected to show? In other words, will resetting be beneficial or detrimental there in comparison to the reset-free process? Mathematically the task is to find the parameter space where $\langle T_R \rangle < \langle T \rangle $. It turns out that for the case of exponential resetting time distribution one can indeed derive a simpler criterion which will guarantee that resetting will be beneficial. Let us start with a region of parameter space such that resetting indeed helps in reducing the MFPT compared to the underlying reset-free process. Mathematically this is ensured if for any small amount of resetting $\delta r \to 0$ one has
\begin{align}
       \langle T_R(\delta r,\lambda,v) \rangle<\langle T\rangle,
\end{align}
where $\langle T\rangle$ is the MFPT of the underlying reset-free process and $ \langle T_R(\delta r,\lambda,v) \rangle$ is given in \eref{mfptexp}. Here we should emphasize that the above condition is sufficient for resetting to be helpful but not a necessary one. An expansion of the LHS of the above equation up to $\mathcal{O}(\delta r)$ and setting the first order correction term to be less than zero, leads to the following interesting relation (detailed derivation in \aref{appc})
\begin{align}
    CV^2>  \frac{2}{\langle T \rangle} \overline{\langle t^{ret}(x) \rangle}+2\overline{\epsilon^{ret}_O(x)}-1, \label{cv-1d}
\end{align}
where $CV$ is the coefficient of variation of the underlying reset-free process defined as
\begin{align}
    CV=\frac{\sqrt{\langle T^2 \rangle -\langle T \rangle^2}}{\langle T \rangle}.
\end{align}
This is a measure of the relative fluctuation of the first passage time of the underlying reset-free process. The overbar $\overline{(...)}$ in \eref{cv-1d} implies a average has to be taken with  $G_0(x)$ so that $\overline{f(x)}=\int_{-\infty}^Ldx~G_0(x)f(x)$. The quantity $G_0(x)$ is defined as
\begin{align}
    G_0(x) = \frac{1}{\langle T \rangle}\int_0^\infty dt G(x,t),
\end{align}
which is normalized by construction so that $\int_{-\infty}^LG_0(x)dx=1$. The condition in \eref{cv-1d} is not limited to the drift-diffusive system considered here, but applies to any arbitrary first passage system in 1-d. For the drift-diffusive system considered in this paper, the mean and the second moment are given, respectively, by $\langle T \rangle=L/v$ and $\langle T^2 \rangle=\frac{L (2 D+L v)}{ v^3}$ (which can be found from Taylor series expansion of \eref{tl-exp}). The $CV$  thus becomes
\begin{align}
    CV=\sqrt{\frac{2D}{Lv}}=\frac{1}{\sqrt{\text{Pe}}}, \label{cv-dd}
\end{align}
where $\text{Pe}=\frac{Lv}{2D}=\frac{\overline{v}}{2}$ is the so-called Peclet number.  It is simply the ratio of the diffusion and drift time scales. \eref{cv-1d} is quite useful since without any prior knowledge of the resetting rates one can infer from here if resetting will be helpful. From here one can also recover the $CV$ criterion for instantaneous and deterministic return protocols obtained in previous studies as shown below.

\textbf{Instantaneous return limit -} In the instantaneous return limit the particle is guaranteed to reach the origin so that $\epsilon_O(x)=1$. On top of that it returns instantaneously so that $\langle t^{ret}(x) \rangle=0$. Hence for instantaneous return, we have
\begin{align}
    CV>1,
\end{align}
a classic result that was derived earlier in \cite{pal_first_2017,pal2019landau,pal2023random} valid for any arbitrary FP process. This says that, with only information about the mean and fluctuation of the underlying process in hand, one can deduce the utility of resetting.   For the drift-diffusive system particularly this relation implies from \eref{cv-dd}
\begin{align}
    \text{Pe}<1 \implies \overline{v}<2,
\end{align}
for resetting to be beneficial \cite{ray2019peclet}.
%%%%%%%%%%%%%%%%%%%%%%%%%%%%%%%%%%%%%%%%%%%%%%%%%%%%%%%%%%%%%%%%%%%%%%%%%%%%%%%%%%%%%%%%%%%%%%%%%%%%%%%%%%%%%%%%%%%%%%%%%%%%%%%%%%%%%%%%%%%%%%%%%%%%%%%%%%%%%%%%%%%%%%%%%%%%%%%%%%%%%%%%%%%%%%%%%%%%%%%%%%%%%%%%%%%%%%%%
\begin{figure}
    \centering
    \includegraphics[width=8cm]{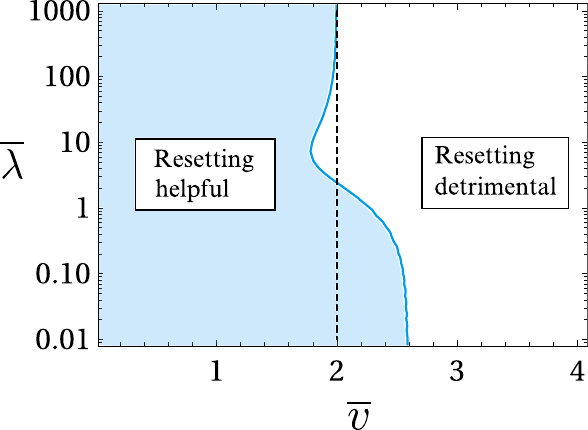}
    \caption{Phase space obtained from the $CV$ criterion in \eref{cv-1d} for the drift-diffusive system with return via $U_{\text{eff}}(x)$. The blue-shaded region is where resetting will help to reduce the MFPT beyond the underlying process. The black dashed line at $\overline{v}=2$ refers to the phase boundary in the instantaneous limit ($\overline{\lambda}\to\infty$).}
    \label{fig3}
\end{figure}
%%%%%%%%%%%%%%%%%%%%%%%%%%%%%%%%%%%%%%%%%%%%%%%%%%%%%%%%%%%%%%%%%%%%%%%%%%%%%%%%%%%%%%%%%%%%%%%%%%%%%%%%%%%%%%%%%%%%%%%%%%%%%%%%%%%%%%%%%%%%%%%%%%%%%%%%%%%%%%%%%%%%%%%%%%%%%%%%%%%%%%%%%%%%%%%%%%%%%%%%%%%%%%%%%%%%%%%%%%

\textbf{Deterministic return limit - } From \eref{cv-1d} it is also possible to obtain the criterion when the return is deterministic. In particular, suppose the particle consumes a finite deterministic time $\tau^{ret}(x)$ to return to the origin starting from $x$.  Note that, in this case also as the particle is guaranteed to reach the origin thus we have $\epsilon_O(x)=1$. Using \eref{cv-1d} one then obtains the criterion for deterministic return case as
\begin{align}
    CV^2>1+\frac{2 \overline{\tau^{ret}(x)}}{\langle T \rangle},
\end{align}
as was obtained in \cite{pal_search_2020}.

\textbf{Stochastic return case -} We now come back to our problem where the particle is taken back to the origin stochastically facilitated by an external potential $U(x)$. One can find the RHS of \eref{cv-1d} exactly by substituting results from \eref{t2-1d}-(\ref{t1-1d}) and \eref{tl-exp}-(\ref{prp_exp}). The resulting expression is quite cumbersome which we provide in the 
\cite{key}. However one can generate the phase space spanned over $\overline{\lambda}-\overline{v}$ plane as shown in \fref{fig3}. If the parameters are chosen from the blue-shaded region, then resetting is bound to decrease the MFPT of this process further below the underlying drift-diffusive limit \textit{i.e.} $1/\overline{v}$ (see \eref{und-process}). The black dashed line ( $\text{Pe}=1$ or $\overline{v}=2$ )represents the phase boundary for the instantaneous resetting limit as obtained previously. Note that, keeping a fixed value of $\overline{\lambda}$ if one gradually increases $\overline{v}$, then at some finite value $\overline{v}$ one crosses from the region where resetting helps to where resetting does not. This behavioural transition is known as \textit{restart transition} and is of second order for our system. A more detailed discussion on restart transition is provided in the \aref{appd}. 

The physical origin behind these two distinct regions can be understood as follows. For very low values of the drift $\overline{v}$ the particle can disperse very far from the origin. These diverging trajectories in the direction opposite to the target can lead to a very high value of the MFPT. Return (any finite value of $\overline{\lambda}>0$) in those cases prohibits the particle from wandering off too far away from the target consequently reducing the MFPT. In contrast, when the drift $\overline{v}$ is sufficiently high, then the MFPT of the underlying process is itself very short. The diverging trajectories as mentioned previously are already very rare in this case. Any kind of resetting in this case would only hinder the trajectories which were supposed to find the target in a shorter time span and cause a delay in the process completion.

An important observation emanating out of \fref{fig3} is the existence of the blue-shaded region even above $\overline{v}=2$. Above this cut-off, instantaneous return can not facilitate the process beyond the underlying one. However, stochastic return turns out to be beneficial here. We shall come back to this discussion again in Sec. \ref{seciv}. In the next subsection, we first try to quantify the criterion when stochastic return is better than instantaneous return for a given rate $\overline{r}$.
\begin{figure*}
    \centering
    \includegraphics[width=16cm]{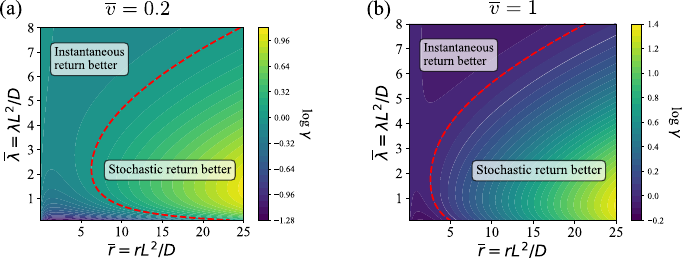}
    \caption{The parameter space where resetting with stochastic return can facilitate the drift-diffusive search process over the instantaneous return, as generated from the SR criterion in \eref{sr-exp-dd}. The colour density denotes the magnitude (log) of the speed-up parameter $\gamma$ as defined in \eref{speedup-exp}. The higher it is the more benefits one can gain with the stochastic return over the classical instantaneous return. The red dashed line is the separatrix which is obtained by setting \eref{sr-exp-dd} to an equality {\it i.e.,} $\gamma=1$. }
    \label{fig5}
\end{figure*}

\subsection{Speed up over instantaneous resetting- the \textit{SR criterion}}
\label{sec3b}
The stochastic mode of return has two major impacts on the search time of the system compared to the classical instantaneous return. The first one is that, here the searcher consumes a finite amount of time to return to the origin. In contrast, in the instantaneous return case, the searcher takes zero time to return. This is indeed what we observe in \fref{fig2}(a) for $\overline{r}=0.1,1$, where any finite $\overline{\lambda}$ yields a higher value of MFPT than that at $\overline{\lambda}\to\infty$ (instantaneous limit). However, since the searcher is returning in a stochastic manner, it is possible once in a while, that it actually finds the target while returning home (origin) and the search process is completed. In the instantaneous return case, however, the searcher always returns to the origin in no time and the whole search process is again restarted. Thus in this case one may see certain advantages in the stochastic return motion. In \fref{fig2}(a) for $\overline{r}=5,10,20$ note that there exists a range of $\overline{\lambda}$ values where MFPT is lowered than that of the instantaneous limit. This behaviour is also observed in all the curves in \fref{fig2}(c). Evidently, there exists a trade-off between the return time and the target search probability. To understand this trade-off quantitatively we simply compare the MFPT for these two modes of return and look at the condition when  MFPT with stochastic return is less than that obtained for classical instantaneous return \textit{i.e.} 
\begin{align}
    \langle T_R \rangle < \langle T^{inst}_R \rangle\label{cond},
\end{align}
where $\langle T^{inst}_R \rangle$ is the MFPT with instantaneous return. One can find $\langle T^{inst}_R \rangle$ from \eref{mfpt-1d} by noting that the return times $\langle t_1(x) \rangle,\langle t_2(x) \rangle$ are zero here as the particle reaches the origin instantaneously. On top of that, as the particle reaches the origin for sure, the splitting probability to the origin is unity \textit{i.e.} $\epsilon_O(x)=1$. Combining both these results we obtain
\begin{align}
    \langle T^{inst}_R \rangle=\frac{\langle min(T,R) \rangle}{Pr(T<R)},
    \label{inst}
\end{align}
where, $Pr(T<R)=1-\int_{-\infty}^Ldx~\widetilde{G}_R(x)$ quantifies the probability that the first passage occurs before any resetting event takes place. The same result was obtained earlier in \cite{pal_first_2017}. For the case when resetting times are exponentially distributed, this takes the form (using \eref{mintr})
\begin{align}
    \langle T^{inst}_R \rangle=\frac{1-\widetilde{T}(r)}{r\widetilde{T}(r)},
\end{align}
where $\widetilde{T}(r)$ is given by \eref{tl-exp}. Note that $\langle T^{inst}_R \rangle=\frac{L^2}{D} \langle \tau(\overline{r},\overline{v}) \rangle_{\text{inst}}$, which can be also be found from \eref{inst-dd} for the drift-diffusive system considered.

Starting from \eref{cond} a general criterion was derived in \cite{biswas2023stochasticity} to identify the phase space regime where stochastic return could be more advantageous. The resulting criterion in 1-d for exponential resetting takes the following form \cite{biswas2023stochasticity}
\begin{small}
\begin{align}
    \mathcal{T}\coloneqq \frac{\int_{-\infty}^Ldx\widetilde{G}(x,r)\langle t^{ret}(x) \rangle}{\int_{0}^Ldx\widetilde{G}(x,r) \epsilon_L(x)}<\langle T^{inst}_R \rangle, 
     \label{sr-exp}
\end{align}
\end{small}where $\epsilon_L(x)$ is the probability of reaching the target during return motion from $x>0$ (the exact expression given in \eref{el}). The quantity $ \mathcal{T}$ quantifies the trade-off between the return times and the target search probability of the stochastic return motion. In particular, the numerator of $ \mathcal{T}$ (in \eref{sr-exp}) takes care of the return times. The higher the return times, the higher the $\mathcal{T}$, adverse to satisfying the criterion in \eref{sr-exp}. The denominator takes care of the probability that the target is found while returning. Thus higher it is the lower becomes the $\mathcal{T}$, which aids in satisfying the criterion (\ref{sr-exp}).  Both the RHS and the LHS of the above inequality can again be written in a dimensionless form to obtain the criterion 
\begin{align}
   \overline{\mathcal{T}}<\langle \tau(\overline{r},\overline{v}) \rangle_{\text{inst}}. \label{sr-exp-dd}
\end{align}
 where we denote the dimensionless quantity $\overline{\mathcal{T}}=\frac{D}{L^2}  \mathcal{T}$. To further quantify the amount of speed-up gained with stochastic return over instantaneous return we define the speed-up parameter 
\begin{align}
    \gamma=\frac{\langle T^{inst}_R \rangle}{ \langle T_R \rangle}=\frac{\langle \tau(\overline{r},\overline{v}) \rangle_{\text{inst}}}{\langle \tau(\overline{r},\overline{\lambda},\overline{v}) \rangle}. \label{speedup-exp}
\end{align}
Clearly, when $\gamma>1$ stochastic return wins. The magnitude of gamma is an indicator of how much gain can be achieved with stochastic return. The separatrix between the two regimes where stochastic or instantaneous return is better can be obtained by setting \eref{sr-exp} to equality \textit{i.e.} from $\gamma=1$.  The LHS of \eref{sr-exp} can be exactly computed by substituting the elements from  \eref{t2-1d}-(\ref{t1-1d}), and \eref{prp_exp}.
Again we do not provide the functional form here due to its complex structure which is provided in \cite{key}. 
%%%%%%%%%%%%%%%%%%%%%%%%%%%%%%%%%%%%%%%%%%%%%%%%%%%%%%%%%%%%%%%%%%%%%%%%%%%%%%%%%%%%%%%%%%%%%%%%%%%%%%%%%%%%%%%%%%%%%%%%%%%%%%%%%%%%%%%%%%%%%%%%%%%%%%%%%%%%%%%%%%%%%%%%%%%%%%%%%%%%%%%%%%%%%%%%%%%%%%%%%%%%%%%%%%
\begin{figure*}
    \centering
    \includegraphics[width=16cm]{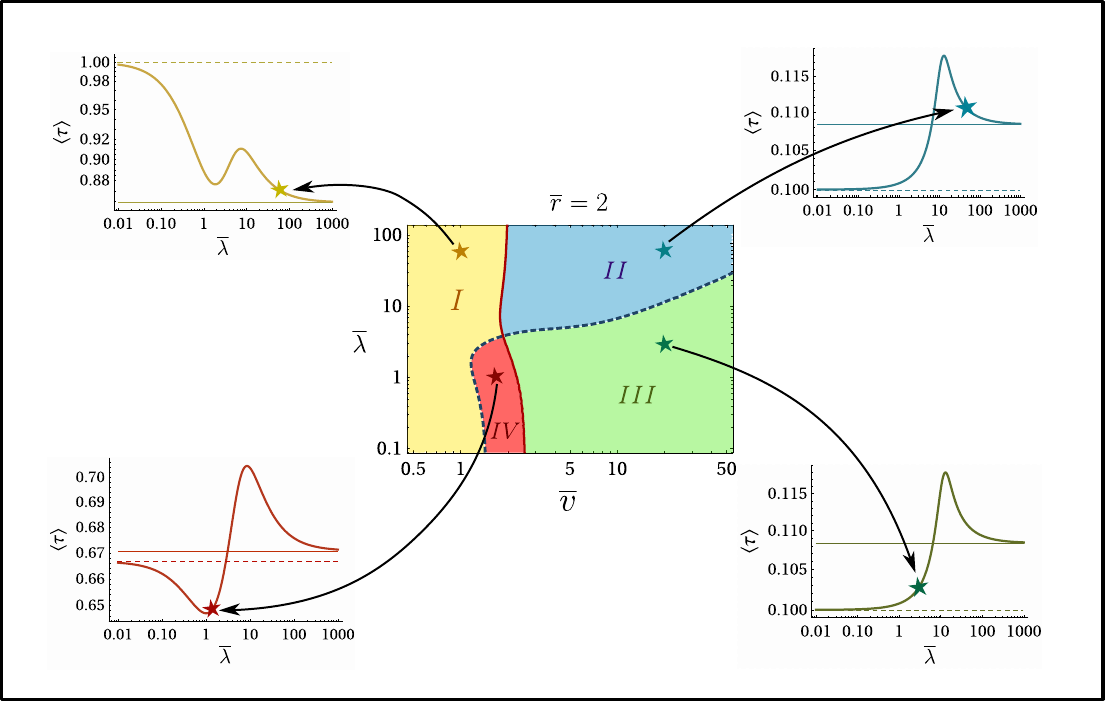}
    \caption{A unified phase diagram for the drift-diffusive search process for exponential resetting with rate $\overline{r}=2$. Different regions in this parameter space represent the regions where either of the criteria in \eref{cv-1d} and \eref{sr-exp-dd} are satisfied. The most important is region IV (the red-shaded region) where both the \textit{CV} and SR criterion is satisfied and stochastic return facilitates the process beyond both the underlying process and also that of resetting with instantaneous return. The behaviour of the MFPT for a sample point (star) in each of these regions is shown in the corners. In these plots, the horizontal solid lines represent the MFPT that with instantaneous return and the dashed horizontal line represents the MFPT of the underlying reset-free process.}
    % \caption{\textcolor{red}{Arup, let's try to maintain the following: Use Pe for the intrinsic drift part while for the $|x|$ potential, use the term potential strength or something similar. Do not bring Pe there for it can create unnecessary fuzz. You can also think alternatively. }}
    \label{fig6}
\end{figure*}
%%%%%%%%%%%%%%%%%%%%%%%%%%%%%%%%%%%%%%%%%%%%%%%%%%%%%%%%%%%%%%%%%%%%%%%%%%%%%%%%%%%%%%%%%%%%%%%%%%%%%%%%%%%%%%%%%%%%%%%%%%%%%%%%%%%%%%%%%%%%%%%%%%%%%%%%%%%%%%%%%%%%%%%%%%%%%%%%%%%%%%%%%%%%%%%%%%%%%%%%%%%%%%%%%%%%%

\fref{fig5} shows the parameter space ($\overline{r}-\overline{\lambda}$) generated by the criterion in \eref{sr-exp-dd} for different values of $\overline{v}$. The colour gradient represents the value of the speed-up parameter $\gamma$. The region right to the separatrix (the red dashed line) is where criterion \eref{sr-exp-dd} is satisfied and stochastic return diminishes the MFPT to a lower value than that obtained through the instantaneous return protocol. Note that the line $\gamma=1$ moves upwards as $\overline{v}$ is increased. This means for a higher value of the drift, the regime where the efficacy of the stochastic return can be realized, gets broadened.

Let us now try to understand the physical origin behind such behaviour of the MFPT. First focus on the regime where $\overline{v}$ is low. In this limit, it is evident from both \fref{fig2}(a) and \fref{fig5}(a) that, instantaneous return always wins for relatively lower values of the resetting rate. As discussed in section \ref{sec3a}, for relatively lower values of $\overline{v}$ the particle may hover far away from the target and consume excessive time to find the target. Resetting cuts off these detrimental trajectories by bringing them back to the origin. However, the crucial point to note here is that stochastic return in this case incurs a huge return time in bringing back the particles from a far away distance. On the other hand, instantaneous return causes them to get back to the origin instantaneously, inducing no extra time. Thus in the limit of small $\overline{v}$ and $\overline{r}$, MFPT of stochastic return is higher than that of instantaneous return. When the resetting rate becomes sufficiently high (irrespective of the value of $\overline{v}$) instantaneous resetting becomes detrimental as it does not allow the particle to even get close to the target. On the other hand for the stochastic return case, the potential is turned on so frequently that the particle essentially moves under the return potential $U_{\text{eff}}(x,t)$ all the time. Evidently, the particle has a chance of reaching the target while moving inside the potential. Consequently, the MFPT is much lower compared to the instantaneous return value.

\subsection{A unified phase diagram}
\label{sec4}

Till now we have found parameter space associated with the two regions, $\langle T_R \rangle < \langle T \rangle$ which is found from the \textit{CV} criterion and $ \langle T_R \rangle < \langle T^{inst}_R \rangle$ which is found from the SR criterion.   Now we are in a position to probe a more intricate aspect: For a given value of $\overline{r}$, can stochastic return do better than both the underlying search process and that with instantaneous return protocol, simultaneously? This is equivalent to finding the parameter space where $\langle T_R \rangle < \langle T \rangle<\langle T^{inst}_R \rangle$ or $\langle T_R \rangle < \langle T^{inst}_R \rangle<\langle T \rangle$.  One can guess this will be the case when both the conditions in \eref{cv-1d} and \eref{sr-exp} are satisfied. To verify this we plot the phase boundary in the $\overline{v}-\overline{\lambda}$ (for a fixed $\overline{r}=2$) plane obtained from these conditions in the central image of \fref{fig6}. The region left to the solid (red) separatrix is where the $CV$ criterion is satisfied. On the other hand, the region below the dashed separatrix is where the SR criterion is satisfied. This immediately presents us with 4 distinct phases in the parameter space. We mark them from I-IV and discuss each of them below in detail.

\textbf{Region I: }In this regime (yellow shaded) $CV$ criterion is satisfied although the SR criterion is not. Consequently, stochastic return is supposed to expedite the search process over the underlying process. However, it can not supersede the instantaneous return protocol. To verify, we pick an arbitrary point from this region at $(\overline{v},\overline{\lambda})\equiv (1,60)$ (the yellow star) and mark its associated MFPT in the plot at the top left corner of \fref{fig6}. The dashed horizontal line shows the MFPT corresponding to the underlying reset-free process and the MFPT associated to resetting with instantaneous return protocol is shown by the solid horizontal line. We can clearly see the MFPT at the yellow star lies above the dashed line (instantaneous return) and below that of the solid line (reset-free process), as expected.

\textbf{Region II: }Here (blue shaded) none of the criteria is satisfied resulting in stochastic return being the most detrimental protocol. For a generic point in this space $(\overline{v},\overline{\lambda})\equiv (20,60)$ (blue star) we see from the MFPT plot at the top right corner of \fref{fig6} that it lies well above both the horizontal solid (instantaneous return) and dashed line (underlying process). 

\textbf{Region III: }This is where (green shaded) only the SR criterion holds but the $CV$ criterion does not.
Evidently, stochastic return is a better alternative than the classical instantaneous return. Although, in this space resetting with stochastic return can not expedite the search process over the underlying process. The same is shown for a point in this space $(\overline{v},\overline{\lambda})\equiv (20,2)$ (green star) which lies below the solid horizontal line (instantaneous return) but above the dashed line (underlying process) in the rightmost plot at the bottom of \fref{fig6}.

\textbf{Region IV: }It is the most noteworthy region (red-shaded) as the supremacy of stochastic return is more pronounced here. Both the SR and $CV$ criteria are satisfied in this space. \textit{Thus stochastic return turns out to be the best way to reduce the search time of the underlying process and also that obtained from instantaneous return protocol}. This is also observed by taking a sample point in this phase space $(\overline{v},\overline{\lambda})\equiv (1.5,1)$ (the red star). In the MFPT vs $\overline{\lambda}$ plot in the bottom left corner of \fref{fig6} we see it lies well below both the solid horizontal line (MFPT due to instantaneous return) and the dashed horizontal line (MFPT of the underlying reset-free process). Moreover, at this point note that instantaneous return does actually increase the mean search time than the underlying process.

To summarize, we mark stochastic return as one of the strategies with which one can expedite a search process even in cases where instantaneous return fails to do so. Note that the phase space as shown in \fref{fig6} is only for the fixed value of $\overline{r}=2$. However, with changing $\overline{r}$ the phase boundaries will change its position and the regions marked by (I-IV) will change.

\section{Speed-up beyond optimal instantaneous return}
\label{seciv}
Let us now take a step backwards and look at the motivation behind restarting a first passage process. The central role of inducing resetting to a search process was to increase the search efficiency \textit{i.e.} to minimize the MFPT. Performing instantaneous resetting on the original process is the first protocol which was seen to mitigate the diverging trajectories going away from the target and thus expediting the search process \cite{evans_diffusion_2011,evans_stochastic_2020,pal_first_2017}. However, when the reset-free process itself has a finite mean completion time $\langle T \rangle$, then performing instantaneous resetting may not always reduce the mean completion time. For example, performing instantaneous resetting too frequently is definitely detrimental. In such a situation, stochastic return may turn out to be helpful in improving the search process.  Due to the finite target-finding ability of the searcher during the return phase stochastic return gains the upper hand there. Despite the utility of stochastic return for a given resetting rate and at places where instantaneous return fails, we are still left to wonder:  Can stochastic return reduce the MFPT of the process even below the \textit{lowest} MFPT that can be obtained with instantaneous return protocol?

% From \fref{fig5} (using criterion \eref{sr-exp}) we show a parameter space where significant advantage can be gained leveraging non-instantaneous stochastic return modulated by a potential trap. Due to the finite target-finding ability of the searcher during the return phase stochastic return gains the upper hand there. But we are still left to wonder whether it can improve over the underlying process as well. This answer is given by the $CV$ criterion as in \eref{cv-1d}. We again find a parameter space (\fref{fig3}) where this criterion is satisfied and resetting with stochastic return is guaranteed to expedite the search process. The discussion in the last section provides us with a parameter space where stochastic return can be better than instantaneous return. However, one can further ask:

\begin{figure*}
    \centering
    \includegraphics[width=17cm]{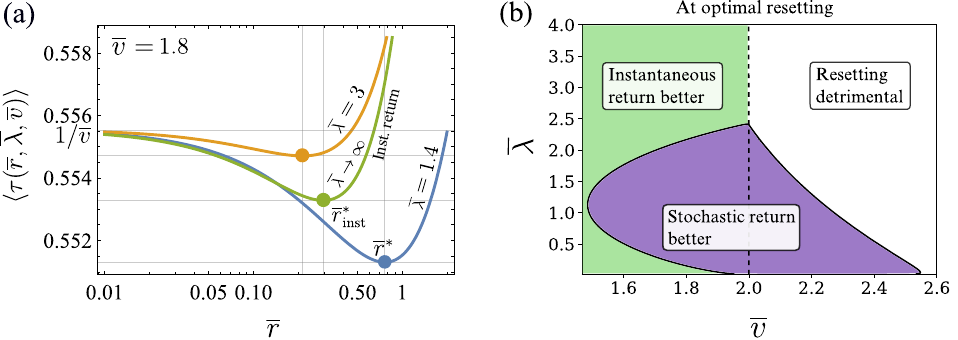}
    \caption{(a) MFPT with stochastic return \textit{i.e.} $\langle \tau(\overline{r},\overline{\lambda},\overline{v}) \rangle$ (blue and orange curves) and instantaneous return \textit{i.e.} $ \langle \tau(\overline{r},\overline{v}) \rangle_{\text{inst}}=\langle \tau(\overline{r},\overline{\lambda} \to \infty,\overline{v}) \rangle$ (green curve) are plotted with respect to the resetting rate $\overline{r}$, for a fixed drift velocity $\overline{v}=1.8$. Remarkably, for $\overline{\lambda}=1.4$ we find that optimal stochastic return can lower the MFPT (the blue circle) even beyond the optimal instantaneous return (the green circle). (b) Phase space showing regions where stochastic (or instantaneous) return is better at the optimal resetting rate (which is different in general for both the protocols). The blue shaded region (obtained by setting the speed-up parameter $\gamma^*$ as in \eref{su-opt}, to be greater than unity) is where stochastic return performs better than the optimal instantaneous return. }
    \label{fig66}
\end{figure*}

\subsection{MFPT at optimal resetting rate}
Let us elaborate more. For a fixed value of drift $\overline{v}$, one observes from \fref{fig2}(b) that for a particular value of the resetting rate $\overline{r}$ the MFPT is lowest. This resetting rate is called optimal resetting rate (ORR) denoted by $\overline{r}^*$. Mathematically this can be found from the equation 
\begin{align}
   \left. \frac{\partial  \langle \tau(\overline{r},\overline{\lambda},\overline{v}) \rangle}{\partial \overline{r}}\right|_{\overline{r}=\overline{r}^*}=0.\label{orr1}
\end{align}
For a fixed value of $\overline{\lambda},\overline{v}$ the lowest MFPT that can be obtained through stochastic return is $\langle \tau(\overline{r}^*,\overline{\lambda},\overline{v}) \rangle$. Note that ORR $\overline{r}^*$ is a function of $\overline{\lambda},\overline{v}$ in general. In the limit $\overline{\lambda}\to\infty$ we obtain the ORR for the instantaneous case as
\begin{align}
    \overline{r}^*_{\text{inst}}=\overline{r}^*(\overline{\lambda}\to\infty,\overline{v}),
\end{align}
which can also be found from \eref{inst-dd} as
\begin{align}
   \left. \frac{\partial   \langle \tau(\overline{r},\overline{v}) \rangle_{\text{inst}}}{\partial \overline{r}}\right|_{\overline{r}=\overline{r}^*_{\text{inst}}}=0.\label{orr-inst}
\end{align}
% Given a fixed value of $\overline{v}$, we ask, can the MFPT with at optimality with stochastic return i.e. $\langle \tau(\overline{r}^*,\overline{\lambda},\overline{v}) \rangle$ can be lowered than the MFPT at the optimal instantaneous return \textit{i.e.} $ \langle \tau(\overline{r}^*_{\text{inst}},\overline{v}) \rangle_{\text{inst}}$? As the 
Thus for a fixed value of $\overline{v}$ the lowest MFPT that can be obtained through instantaneous resetting is $ \langle \tau(\overline{r}^*_{\text{inst}},\overline{v}) \rangle_{\text{inst}}$. However, keeping $\overline{v}$ fixed one can still vary the potential strength $\overline{\lambda}$ for the stochastic return to find the $r^*$ associated with each value of $\lambda$. Can we find a parameters space in $\overline{\lambda},\overline{v}$ where MFPT at this ORR can be compared?

We recall from  Sec. \ref{sec3a} that for $\overline{v}>2$ instantaneous return fails to expedite the underlying process. Consequently, the $r^*_{\text{inst}}$ is exactly zero there (see \aref{appd} for more discussion). On the other hand, a careful look at \fref{fig3} reveals that there is a certain blue shaded region even at $\overline{v}>2$. This is where stochastic return is actually able to enhance the search process compared to the underlying process, in spite of instantaneous return failing to do so. Evidently, $r^*$ is non-zero in this region. Interestingly, there also exists a set of values of $\overline{v}$ (which is less than 2) where despite having a finite $r^*_{\text{inst}}$, stochastic return at $r^*$ can perform better. As an illustrative example look at \fref{fig66}(a) where we plot the MFPT for both the protocols with respect to $\overline{r}$ keeping $\overline{v}=1.8$. Note that the blue curve, corresponding to the MFPT with stochastic return for $\overline{\lambda}=1.4$, shows a minima at $\overline{r}^*$. The MFPT at this minima is lower than the lowest MFPT obtained at optimal instantaneous return (the green curve) at $\overline{r}^*_{\text{inst}}$. Although, for $\overline{\lambda}=3$ (the orange curve) the lowest MFPT obtained with stochastic return is higher than that with instantaneous return. In a nutshell, we have a set of parameters in terms of $(\overline{v},\overline{\lambda})$ where stochastic return at optimal resetting can indeed outperform the optimal instantaneous resetting. To quantify this region we calculate the speed-up parameter $\gamma^*$ at the optimal resetting rate, defined as
\begin{align}
    \gamma^*=\frac{\langle \tau(\overline{r}^*_{\text{inst}},\overline{v}) \rangle_{\text{inst}}}{\langle \tau(\overline{r}^*,\overline{\lambda},\overline{v}) \rangle}. \label{su-opt}
\end{align} 
The region $\gamma^*>1$ is where optimal stochastic return better than optimal instantaneous return, shown by the blue shaded region in \fref{fig66}(b). In the green shaded region ($\gamma^*<1$), instantaneous return prevails. Whereas, in the white region underlying process already performs at its best and any kind of resetting is detrimental here.

\subsection{A possible physical explanation of such behaviour} 
We now make an attempt to understand such speed-up over optimal instantaneous resetting from the behaviour of the process on a trajectory level. We first list the major effects imposed in the search process by stochastic return motion.
\begin{enumerate}[label=(\alph*)]
    \item In the region $x<0$, the effective drift $\lambda_-=\lambda+v$ is always stronger than the underlying drift velocity $v$, which aids in cutting the diverging trajectories in the $x<0$ region appearing in the underlying reset-free process.
    \item In the region $x>0$, the drift $\lambda_+=\lambda-v$ acts in opposite direction of the target when $\lambda>v$ and reduces the net drift towards the target when $\lambda<v$. In both scenarios, the search is hampered compared to the underlying process.
\item On the other hand if we compare to the instantaneous resetting, trajectories return to the origin in zero time during the return phase, adding no additional return time to the MFPT.
\item At $x>0$, however, some trajectories reach the target during return phase due to its stochastic dynamics. Whereas, for the instantaneous return the particle reaches the origin in zero time and thus no target search is possible during return. In this case, stochastic return turns out to be advantageous over the instantaneous return for some choices of parameters.
\end{enumerate}
It is the complex interplay amongst the above-mentioned effects that come into picture for any given set of parameters, which determines whether stochastic or instantaneous return will be the better policy. Let us now explore them taking different limits of the drift $\overline{v}$.

\textit{1) Very high $\overline{v}\gg 2$}: From \fref{fig66}(b) (or \fref{fig3}) it is clear that for very high values of the drift velocity $\overline{v}$, neither of stochastic or instantaneous return can expedite the search process. For very high values of drift towards the target, most of the trajectories are expected to be going towards the target in a very short time. Bringing back the particle to the origin (instantaneously or even turning on the return potential) only inhibits these trajectories, which leads to higher values of the MFPT. Thus any kind of resetting is detrimental in those cases.

\textit{2) Intermediate value of $\overline{v} \gtrsim 2$}: For intermediate values of $\overline{v}\gtrsim 2$, from \fref{fig66}(b) we find there exists a region (shown by the violet shaded region above the dashed vertical line at $\overline{v}=2$) where instantaneous return does not help in expediting the search process but stochastic return protocol does. This might be due to the following reasons. The MFPT of the underlying process increases here from the previous case as the number of trajectories going away from the target is comparatively higher here. Yet, most of the trajectories again go towards the target and restarting them instantaneously only delays the process. Clearly, instantaneous return again turns out to be detrimental. On the other hand, for the stochastic return case when the potential is turned on (at suitably chosen value of $\overline{\lambda}$, which is not very high) the particle does have a finite chance to reach the target rather than returning to the origin. This target search probability suppresses the effect due to finite return time from trajectories going away from the target. So effect (d) is more prominent here compared to (c). Alternatively, now the trajectories going away from the target $(x<0)$ experience a higher value of net drift towards the origin (as $\lambda_->\overline{v}$) which obviously helps in preventing trajectories from wandering off for a long time compared to the underlying process. Here effect (a) is more pronounced than effect (b). Consequently, stochastic return turns out to be a better strategy here compared to the underlying process. However, for $\overline{\lambda}\gg 1$ the target search probability during return is significantly reduced and those trajectories contribute to a higher mean return times. As a result effect (c) now becomes stronger to make the stochastic return protocol inefficient compared to the instantaneous return protocol.

\textit{3) Intermediate value of $\overline{v} \lesssim 2$}: The blue region left to the dashed line \fref{fig66} is where optimal instantaneous return helps in reducing the MFPT compared to the underlying process. However, stochastic return protocol at optimal values of the resetting rate and suitable $\overline{\lambda}$ can still minimize the MFPT beyond the optimal instantaneous return here. The underlying physical reason can again be explained similarly to the previous case. When drift is low enough ($\overline{v}<2$), a sufficient number of diverging trajectories are affected by the optimal instantaneous return reducing the MFPT. However, the finite target search probability during stochastic return fashion (for suitably chosen values of $\overline{\lambda}$) still dominates over effect (d). So stochastic return is still a better strategy here. Moreover, here we see for comparatively lower values of $\overline{v}$, when $\overline{\lambda}\ll 1$ stochastic return fails to expedite the search compared to the instantaneous return. This is because for low values of $\overline{v}$ the diverging trajectories incur higher return times, so that effect (c) suppresses effect (d). Consequently, the stochastic return becomes detrimental. As previously discussed, for $\overline{\lambda}\gg 1$ the target search probability gets reduced so that (c) is more effective than (d), and stochastic return loses its benefits. 

\textit{4) Very low values of $\overline{v}\ll 2$}: In this case, the number of trajectories going away from the target increases significantly. Evidently, the return times are huge even at optimality for the stochastic return mode. Clearly, instantaneous return wins here.

 % For intermediate values of $\overline{v}$ from \fref{fig3} we find there exists a region (shown by the blue shaded region above the dashed vertical line at $\overline{v}=2$) where instantaneous return does not help in expediting the search process but stochastic return protocol does. From \fref{fig66} it is also observed that this region where stochastic return can be better than optimal instantaneous return also extends even below $\overline{v}=2$.  This might be due to the following reasons. When $\overline{v}$ is sufficiently high, 

\section{Discussion and outlook}
In this paper, we studied the drift-diffusive search process in 1-d under resetting with stochastic return. The return phase is modulated by turning on a linear potential trap at the origin. When the trap strength is sufficiently high, one recovers the limit of resetting with instantaneous return. We particularly focus on two aspects of stochastic return motion: first, whether it can reduce the MFPT beyond the underlying reset-free process and second, whether it can also expedite the search over the classical instantaneous return protocol for a given resetting rate. To answer the former, we derive the $CV$ criterion which guarantees that resetting with stochastic return indeed enhances the search process compared to the underlying process. For the later one, we derive the SR criterion which ensures that the stochastic return protocol does better than the instantaneous return protocol. Then, we combine both these criteria to find a unified parameter space where stochastic return turns out to be a potential strategy to expedite the FP process over both the underlying process and also that with instantaneous return protocol. Finally, we investigate the question: can stochastic return perform better than the instantaneous return at their respective optimality? We, indeed, find a parameter space where despite the existence of a non-zero optimal instantaneous resetting rate, stochastic return is seen to perform better in reducing the search completion time further.

It is indeed remarkable that in spite of an overhead time caused by space-time coupled return protocol, stochastic return does a good job of decreasing the MFPT. This is primarily due to the target-finding ability of the searcher during the return phase. Besides, stochastic return provides a more realistic description of a resetting event. For example, in experiments involving the colloidal particles resetting is implemented by turning on a potential trap at the origin. In those scenarios, there is always a chance that the particle goes uphill and finds the target. Those trajectories are usually filtered out in experiments as they are theoretically hard to take into account. However, we show that this can actually be done and also greatly helps in expediting the search process.

The study of stochastic return in the context of resetting processes is still at its infancy and many generalizations can be proposed. A straightforward extension would be to study the problem in higher dimensions. It would also be interesting to look at its effects when more than one target is present. Apart from that it still remains as an open question: what is the optimal choice for the resetting time distribution $f_R(t)$ so that the mean completion time can be globally minimized? 
%For the case of classical instantaneous return, it has been found that sharp/periodic resetting protocol is the best policy. However, for the stochastic return protocol, this still remains to be answered.  
Finally, we would like to conclude by emphasising that stochastic return motion can indeed be beneficial for other search processes that go beyond diffusion and it would be tempting to know how an analogous SR criterion can be manifested in those systems. With the parameters space derived from those criteria, one can easily navigate between the best choice of search strategy to facilitate the process.

\section{Acknowledgement} AK acknowledges the support of the core research grant CRG/2021/002455
and the MATRICS grant MTR/2021/000350 from the SERB, DST, Government of India. AK also acknowledges support of the Department of Atomic Energy, Government of India, under Project No. RTI4001. AP gratefully acknowledges research support from the Department of Science and Technology, India, SERB Start-up Research Grant Number SRG/2022/000080 and Department of Atomic Energy, Government of India.

\section*{Appendices}
\appendix
\section{Calculations of MFPT and splitting probabilities of drift-diffusive search in 1-d}
\label{appa}
In this section, we shall provide the detailed calculations for the basic observables (as in \eref{t2-1d}-(\ref{t1-1d})) of a 1-d drift-diffusive search process. Note that when the return phase starts from a position $x_0>0$ then the origin acts as a virtual absorbing boundary. This is because, in this case, the return phase is terminated whenever the particle makes a first passage to the origin along with the target at $x=L$.  Let us consider a Brownian particle that diffuses in the interval $x\in [0,L]$. The boundaries at $x=0,L$ are purely absorbing in nature. In addition to its diffusive motion, it experiences drift of magnitude $\lambda_d>0$ towards the boundary at $x=L>0$. The drift $\lambda_d$ can stand form both $\lambda_+=\lambda-v$ (for $x>0$) or $\lambda_-=\lambda+v$ (for $x<0$) as in \eref{ueff}. One can write the associated drift-diffusion equation for the probability density $ G_{ret}(x,t)$ as
\begin{align}
    \frac{\partial G_{ret}(x,t)}{\partial t}-\lambda_d \frac{\partial G_{ret}(x,t)}{\partial x}=D\frac{\partial^2 G_{ret}(x,t)}{\partial x^2}
    \label{dd-fp},
\end{align}
with the initial condition 
\begin{align}
    G_{ret}(x,t=0)=\delta (x-x_0), \label{ic}
\end{align} 
where $x_0$ is the particle's position at $t=0$. The boundary conditions read
\begin{align}
    G_{ret}(x=0,t)=G_{ret}(x=L,t)=0. \label{bc}
\end{align} 
In Laplace space, this equation takes the form
\begin{align}
    D\frac{\partial^2 \widetilde{G}_{ret}(x,s)}{\partial x^2}+\lambda_d \frac{\widetilde{G}_{ret}(x,s)}{\partial x}-s\widetilde{G}_{ret}(x,s)=-\delta (x-x_0), \label{GLS}
\end{align}
where $\widetilde{G}_{ret}(x,s)=\int_0^\infty G_{ret}(x,t)e^{-st}dt$. The above equation can be  solved exactly to get a closed form expression for $\widetilde{G}_{ret}(x,s)$ as
\begin{align}
  & \widetilde{G}_{ret}(x,s)= \nonumber \\
  &-\frac{e^{\frac{\lambda_d  (x_0-x)}{2 D}} \text{csch}\left(mL\right)}{2mD} \Bigg[\theta (x-x_0) \Big(\cosh \left(m(L+x-x_0)\right)\nonumber \\
   &\hspace{4.6cm}-\cosh \left(m(L-x+x_0)\right)\Big)\nonumber \\
   &\hspace{1cm}+ \cosh \left(m(L-x-x_0)\right)-\cosh \left(m(L+x-x_0)\right)\Bigg],
   \label{gls}
\end{align}
where, $m=\frac{\sqrt{\lambda_d^2 + 4Ds}}{2D}$ and $\theta(x)$ is the Heaviside step function. 
\subsection{Mean first passage times}
To find the mean first passage times of the particle we need to find the survival probability which can be computed as
\begin{align}
   Q(t)=\int_{0}^L~dx~G_{ret}(x,t),
\end{align}
which in Laplace space reads
\begin{align}
    \widetilde{Q}(s)=\int_0^{\infty}dt e^{-st} Q(t)=\int_{0}^L \widetilde{G}_{ret}(x,s) dx.
\end{align}
From here the mean first passage time of the particle to reach either of the boundaries is given by
\begin{align}
   \langle t_2(x_0) \rangle&= \lim_{s\to 0}  \widetilde{Q}(s)
   =\frac{L(1-e^{\lambda_d x_0/D}) +x_0(e^{\lambda_d L/D }-1)}{\lambda_d(e^{\lambda_d L/D}-1)}.
   \label{ucmfpt}
\end{align}
With $\lambda_d=\lambda_+=\lambda-v$ we obtain \eref{t2-1d} in the main text \textit{i.e.} the MFPT when the return phase is started from $x_0>0$. However, when the return is started from $x_0<0$, there is no target detection possible and in this case, the required result can be obtained by taking limit $L\to\infty$ in Eq.~\eqref{ucmfpt}, and one finds

\begin{align}
    \langle t_1(x_0) \rangle=\frac{|x_0|}{\lambda_d}, \label{ucmfpt1}
\end{align}
Setting $\lambda_d=\lambda_-=\lambda+v$ one obtains \eref{t1-1d} of the main text.

\subsection{Splitting probabilities}
The probability flux through each of the boundaries in Laplace space is given by \cite{redner2001}
\begin{align}
    j_L(x_0,s)&=-D\frac{\partial  \widetilde{G}_{ret}(x,s)}{\partial x}\bigg|_{x=L}, \label{jl}\\
        j_O(x_0,t)&=D\frac{\partial  \widetilde{G}_{ret}(x,s)}{\partial x}\bigg|_{x=0}.
        \label{jo}
\end{align}
Using these one can obtain the splitting probabilities for each of the targets as
\begin{align}
   \epsilon_L(x_0)=j_L(x_0,s\to 0)&=\frac{1-e^{\lambda_d x_0/D}}{1-e^{\lambda_d L/D}}, \\
    \epsilon_O(x_0)=j_O(x_0,s\to 0)&=1-\epsilon_L(x_0) \nonumber\\
    &=\frac{e^{\lambda_d x_0/D}-e^{\lambda_d L/D}}{1-e^{\lambda_d L/D}},
\end{align}
 as announced \eref{el} of the main text. Here $\lambda_d=\lambda_+=\lambda-v$ \textit{i.e.} when return commences from $x_0>0$. For $x_0<0$, the target detection is not possible and the associated splitting probabilities are simply given by \eref{e-neg}.

\section{Derivation of $\langle min(T,R)\rangle$ in \eref{prz} }
\label{appc1}
Let $Z$  represent the random variable 
\begin{align}
    Z=min(T,R),
\end{align}
with the associated density $f_Z(t)$. We thus have
\begin{align}
    \langle min(T,R)\rangle=\int_0^\infty t f_Z(t) dt. \label{c2}
\end{align}
The density $f_Z(t)$ can be found from the cumulative distribution of $Z$ as
\begin{align}
    f_Z(t)=-\frac{d}{dt}Pr(Z>t). \label{fz}
\end{align}
Now the cumulative distribution $Pr(Z>t)$ can easily be found by noting that the condition $Z=min(T,R)>t$ holds only when both of $T$ and $R$ are simultaneously greater than $t$. That in turn implies
\begin{align}
    P(Z>t)=Pr(T>t)Pr(R>t). \label{pzt}
\end{align}
One can now do an integration by parts of the integral in \eref{c2} to have
\begin{align}
    \langle min(T,R)\rangle&=\int_0^\infty t f_Z(t) dt\nonumber\\
    &=\left[t\int f_Z(t)dt\right]_0^\infty-\int_{0}^\infty dt \left(\int dt f_Z(t)\right).
\end{align}
From \eref{fz} we have $\int f_Z(t)dt=-Pr(Z>t)$ which upon inserting to the above equation we have
\begin{align}
     \langle min(T,R)\rangle&=\lim_{t\to \infty}tPr(Z>t)+\int_0^\infty dtPr(Z>t).
\end{align}
Assuming $Pr(Z>t)$ to decay faster than $t$ as $t\to\infty$ (note that this condition holds trivially for the drift-diffusive system with exponential resetting as both of $Pr(T>t)$ and $Pr(R>t)$ decay exponentially, which can be verified from the respective distributions \eref{fpt-dd} and \eref{exp}), we finally have
\begin{align}
    \langle min(T,R)\rangle=\int_0^\infty dtPr(Z>t),
\end{align}
which upon substituting result form \eref{pzt} yields
\begin{align}
     \langle min(T,R)\rangle=\int_0^\infty dt Pr(T>t)Pr(R>t).
\end{align}
This is the same result as in \eref{prz} of the main text.

\section{First passage time distribution of drift-diffusion process in 1-d}
\label{appb}
In this section, we derive the first passage time distribution of a 1-d drift-diffusive process in presence of a single absorbing boundary at $x=L$, as given in \eref{fpt-dd}. If $f_T(t)$ denotes the first passage time distribution then it is related to the survival probability $Q(t)$ as
\begin{align}
    f_T(t)=-\frac{\partial Q(t)}{\partial t}.
    \label{def:f_T(t)}
\end{align}
where the survival probability $Q(t)$ is given by 
\begin{align}
    Q(t)=\int_{-\infty}^{L}dx~G(x,t)~.
    \label{def:Q(t)}
\end{align}
The drift-diffusive propagator $G(x,t)$ is given by \eref{prp}. Inserting this expression of $G(x,t)$ in Eqs.~(\ref{def:Q(t)}, \ref{def:f_T(t)}) and performing the integral, one can obtain the first passage time distribution as 
\begin{align}
      f_T(t)&=\frac{L e^{-\frac{(L-t v)^2}{4 D t}}}{ \sqrt{4 \pi D t^3} },
\end{align}
as was written in \eref{fpt-dd}.

\section{Derivation of the \textit{CV} criterion for resetting with stochastic return}
\label{appc}
In this section, we shall derive the $CV$ criterion as in \eref{cv-1d}. We start by considering a small resetting rate $\delta r\to 0$ limit of  \eref{mfptexp}. The reason for doing so is that whenever the first order term $\mathcal{O}(\delta r)$ is negative, the MFPT of the reset process is guaranteed to be smaller than that of the underlying process. For reader's convenience, we rewrite \eref{mfptexp} here
\begin{align}
    &  \langle T_R(r,\lambda,v) \rangle =\frac{\frac{1-\widetilde{T}(r)}{r} +r\int_{-\infty}^Ldx\widetilde{G}(x,r)\langle t^{ret}(x) \rangle}{1-r\int_{-\infty}^Ldx~\widetilde{G}(x,r) \epsilon^{ret}_O(x)}. \label{amfpt}
\end{align}
For this let us first expand \eref{amfpt} with respect to $\delta r \to 0$. Note that in this limit we have
\begin{align}
    \widetilde{T}(\delta r \to 0)=1-\delta r\langle T \rangle+\frac{\delta r^2}{2}\langle T^2 \rangle +\mathcal{O}(\delta r^3).
\end{align}
Substituting this in \eref{amfpt} we obtain
\begin{align}
    &\langle T_R(\delta r,\lambda,v) \rangle\nonumber\\
    &=\left(\langle T \rangle -\frac{\delta r}{2} \langle T^2 \rangle+\delta r\int_{-\infty}^Ldx\widetilde{G}(x,0)\langle t^{ret}(x) \rangle \right) \nonumber\\
    &\hspace{1.3cm} \times\left(1+\delta r\int_{-\infty}^Ldx~\widetilde{G}(x,0) \epsilon^{ret}_O(x)\right).
\end{align}
 With slight rearrangement, we obtain
\begin{align}
   &  \langle T_R(\delta r,\lambda,v) \rangle  \nonumber\\
   &=  \langle T \rangle +\delta r \Bigg(-\frac{\langle T^2 \rangle}{2}+\int_{-\infty}^Ldx\widetilde{G}(x,0)\langle t^{ret}(x) \rangle \nonumber\\
   &\hspace{2cm}+\langle T \rangle \int_{-\infty}^Ldx~\widetilde{G}(x,0) \epsilon^{ret}_O(x) \Bigg)  \\ 
   &+\mathcal{O}(\delta r^2). \notag
\end{align}
The first term is the MFPT of the underlying reset-free process. Thus any infinitesimal $\delta r$ is supposed to decrease the MFPT beyond that only when the term under the parentheses is less than zero. In that case, we obtain
\begin{align}
   &\Bigg( -\frac{\langle T^2 \rangle}{2}+\int_{-\infty}^Ldx\widetilde{G}(x,0)\langle t^{ret}(x) \rangle \nonumber\\
   &\hspace{2.6cm}+\langle T \rangle\int_{-\infty}^Ldx~\widetilde{G}(x,0) \epsilon^{ret}_O(x) \Bigg)<0 \nonumber\\
   &\implies\langle T^2 \rangle>2 \int_{-\infty}^Ldx\widetilde{G}(x,0)\langle t^{ret}(x) \rangle\nonumber\\
   &\hspace{3cm}+2\langle T \rangle\int_{-\infty}^Ldx~\widetilde{G}(x,0) \epsilon^{ret}_O(x) \nonumber\\
    &\implies\langle T^2 \rangle-\langle T \rangle^2 >2 \int_{-\infty}^Ldx\widetilde{G}(x,0)\langle t^{ret}(x) \rangle\nonumber\\
   &\hspace{2cm}+2\langle T \rangle\int_{-\infty}^Ldx~\widetilde{G}(x,0) \epsilon^{ret}_O(x)-\langle T \rangle^2\nonumber\\
   &\implies \frac{\langle T^2 \rangle-\langle T \rangle^2}{\langle T \rangle^2}>\frac{2}{\langle T \rangle^2}\int_{-\infty}^Ldx\widetilde{G}(x,0)\langle t^{ret}(x) \rangle \nonumber\\
   & \hspace{3cm}+\frac{2}{\langle T \rangle} \int_{-\infty}^Ldx~\widetilde{G}(x,0)\epsilon^{ret}_O(x)-1.
\end{align}
Noting that that $\widetilde{G}(x,r)=\int_{0}^{\infty}dt~e^{-rt}G(x,t) \implies \widetilde{G}(x,0)=\int_{0}^{\infty}dt~G(x,t) $ we have
\begin{align}
  & \frac{\langle T^2 \rangle-\langle T \rangle^2}{\langle T \rangle^2}>\frac{2}{\langle T \rangle^2}\int_{-\infty}^Ldx\int_{0}^{\infty}dt~G(x,t)\langle t^{ret}(x) \rangle \nonumber\\
  &\hspace{2cm} +\frac{2}{\langle T \rangle} \int_{-\infty}^Ldx~\int_{0}^{\infty}dt~G(x,t)\epsilon^{ret}_O(x)-1 \nonumber\\
  &\implies\frac{\langle T^2 \rangle-\langle T \rangle^2}{\langle T \rangle^2}\nonumber\\
  &\hspace{1.5cm}>\frac{2}{\langle T \rangle}\int_{-\infty}^Ldx \left(\frac{1}{\langle T \rangle}\int_{0}^{\infty}dt~G(x,t)\right)\langle t^{ret}(x) \rangle \nonumber\\
  &\hspace{1.5cm} +2 \int_{-\infty}^Ldx~\left(\frac{1}{\langle T \rangle}\int_{0}^{\infty}dt~G(x,t)\right)\epsilon^{ret}_O(x)-1.
\end{align}
Finally identifying the coefficient of variation ($CV$) as $ CV=\frac{\sqrt{\langle T^2 \rangle -\langle T \rangle^2}}{\langle T \rangle}$ we have the following the $CV$ criterion
\begin{align}
    &CV^2>  \frac{2}{\langle T \rangle} \overline{\langle t^{ret}(x) \rangle}+2\overline{\epsilon^{ret}_O(x)}-1,
\end{align}
 as written in \eref{cv-1d} of the main text. Recall that we use the notation $\overline{f(x)}=\int_{0}^\infty dx f(x) G_0(x)$ with $G_0(x)$ being defined as 
\begin{align}
      G_0(x) = \frac{1}{\langle T \rangle}\int_0^\infty dt G(x,t).
\end{align}
%%%%%%%%%%%%%%%%%%%%%%%%%%%%%%%%%%%%%%%%%%%%%%%%%%%%%%%%%%%%%%%%%%%%%%%%%%%%%%%%%%%%%%%%%%%%%%%%%%%%%%%%%%%%%%%%%%%%%%%%%%%%%%%%%%%%%%%%%%%%%%%%%%%%%%%%%%%%%%%%%%%%%%%%%%%%%%
\begin{figure}
    \centering
    \includegraphics[width=7.85cm]{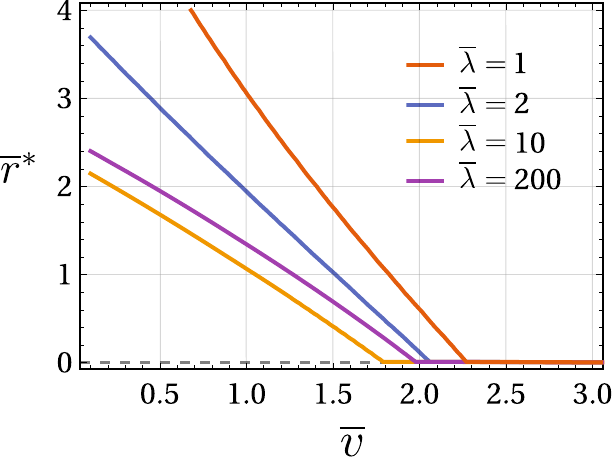}
    \caption{Variation of the optimal resetting rate (ORR) $\overline{r}^*$ with respect to the underlying drift velocity $\overline{v}$ for different values of the potential strength $\overline{\lambda}$. The ORR shows a second-order phase transition with respect to $\overline{v}$. }
    \label{fig4}
\end{figure}
%%%%%%%%%%%%%%%%%%%%%%%%%%%%%%%%%%%%%%%%%%%%%%%%%%%%%%%%%%%%%%%%%%%%%%%%%%%%%%%%%%%%%%%%%%%%%%%%%%%%%%%%%%%%%%%%%%%%%%%%%%%%%%%%%%%%%%%%%%%%%%%%%%%%%%%%%%%%%%%%%%%%%%%%%%%%%%

\section{Restart transition}
\label{appd}
In this section, we investigate the restart transition as briefly mentioned in section \ref{sec2} of the main text. When $\overline{v}$ is low enough (recall \fref{fig2}(b)), note that the MFPT becomes the lowest at some finite value of $\overline{r}=\overline{r}^*$. This resetting rate $\overline{r}^*$ is formally known as the \textit{optimal resetting rate} (ORR). Mathematically, this can be found by setting 
\begin{align}
   \left. \frac{\partial  \langle \tau(\overline{r},\overline{\lambda},\overline{v}) \rangle}{\partial \overline{r}}\right|_{\overline{r}=\overline{r}^*}=0.\label{orr}
\end{align}
A non-zero value of $\overline{r}^*$ ensures that resetting can benefit the search process in lowering the MFPT to its lowest. In contrast, one observes in \fref{fig2}(d), that for a higher value of $\overline{v}$ resetting only increases the MFPT. Consequently, the reset-free process performs at its best and the optimal resetting rate $\overline{r}^*$ is zero. Solving \eref{orr} numerically we can now probe this behavioural transition as depicted in \fref{fig4}. We find that $\overline{r}^*$ undergoes a second-order transition as $\overline{v}$ is gradually increased, for any values of $\overline{\lambda}$. For very high potential strength ($\overline{\lambda}=200$) the transition occurs at $\overline{v}=2$, commensurate with the instantaneous return limit \cite{ray2019peclet}.

\newpage

\bibliography{fpusr}

\end{document}